\documentstyle[12pt,float,psfig]{article}
\restylefloat{figure}
\restylefloat{table}

\advance \topmargin by -\headheight
\advance \topmargin by -\headsep
\evensidemargin \oddsidemargin
\title{Motion of Contact Line of a Crystal Over the Edge of Solid Mask in Epitaxial Lateral Overgrowth}

\author{M. Khenner
\vspace{0.5cm}\\
Department of Mathematics \\
State University of New York at Buffalo \\
Buffalo, NY 14260\\
email: mkhenner@nsm.buffalo.edu}

\renewcommand{\theequation}{\arabic{section}.\arabic{equation}}
\newcommand{\Section}[1]{\setcounter{equation}{0} \section{#1}}
\newcommand{\rf}[1]{(\ref{#1})}
\newcommand{\beq}[1]{ \begin{equation}\label{#1} }
\newcommand{\eeq}{\end{equation} }

\begin{document}
\pagestyle{plain}
\maketitle

\begin{abstract}

\noindent

Mathematical model that allows for direct tracking of the homoepitaxial crystal growth out of
the window etched in the solid, pre-deposited layer on the substrate is described. The growth is governed
by the normal (to the crystal-vapor interface) flux from the vapor phase and by the interface diffusion. 
The model accounts for possibly inhomogeneous energy of the mask surface
and for strong anisotropies of crystal-vapor interfacial energy and kinetic mobility.
Results demonstrate that the motion of the crystal-mask contact line slows down abruptly as radius of curvature 
of the mask edge approaches zero. 
Numerical procedure is suggested to overcome difficulties associated with ill-posedness of the evolution 
problem for the interface with strong energy anisotropy.



\end{abstract}

\Section{Introduction}

\label{Sec1}

The Epitaxial Lateral Overgrowth (ELO) and Selective Area Epitaxy (SAG) are commonly used
to grow micro-scale semiconductor crystals and thin films. 
The micrometer-scale selective growth is well-achievable by
chemical vapor deposition or liquid phase epitaxy \cite{THRUSH}-\cite{Greenspan}; the ELO and SAG can be used also to
grow nanostructures by the molecular beam epitaxy \cite{Lee}-\cite{Mei}.

Recently, the mathematical models  that allow to numerically study the ELO and SAG
from vapor \cite{KBM1} - \cite{KBM3} were introduced. These continuum, geometric crystal growth 
models are formulated as 
free-boundary problems in 
two dimensions
(that is, normal to the substrate/mask and normal to the axis of long stripe openings 
etched in the mask) and, therefore,
they are capable of explicitly accounting for the mask topography and the
interface 
anisotropy. The crystal growth is by the normal  
flux of particles from the vapor phase to the interface; the surface diffusion redistributes the material along the 
interface during the growth. Evaporation-recondensation from both the mask surface and crystal-vapor interface, and
the diffusion of adatoms along the mask surface to the tri-junction crystal-mask-vapor
are accounted for by the models.
The details of the formulation, boundary conditions and numerical method can be found
in \cite{KBM1,KBM2}. In this paper we only address issues related to the mask surface modeling
and the crystal interaction with the edge of the mask (mask corner point). Note that
the crystal growth out of the trench occurs relatively early in the ELO process; however,
our previous modeling suggests \cite{KBM1,KBM2} that the perturbation introduced to the crystal-vapor
interface by the edge of the mask  exists for a long time after the actual interaction of the advancing 
edge of a crystal (contact line) and the mask edge took place. Thus it seems important to understand
the details of the contact line motion over the edge.

In \cite{KBM1,KBM2} the treatment of the contact line motion relies on the assumptions of 
(a) the thermodynamic equilibrium
at the tri-junction (indeed, SAG and ELO are equilibrium processes), (b) the mask surface as the
rectangular, macroscopic step, and (c) the mask as the energetically homogeneous surface: 
the mask-vapor surface energy $\gamma_{mv} = const.$ The assumptions (b) and (c) 
limit the ability of any model to describe the overgrowth of the crystal line from the vertical onto
the horizontal part of the mask. Indeed, (a) implies that the equilibrium contact angle $\phi$ 
is given by the equation
\begin{equation}
\gamma_{mv}-\gamma_{cm} - \gamma_{cv}\cos{\phi} + \frac{\partial \gamma_{cv}}{\partial \phi}\sin{\phi} = 0,
\label{1.3}
\end{equation}
where $\gamma_{cm}$ and $\gamma_{cv}$ are the energies of the crystal-mask and crystal-vapor interfaces, respectively
(see Appendix A).
Then, (c) together with the natural assumptions of constant (at the tri-junction) $\gamma_{cm}$ and
$\gamma_{cv}$ imply that $\phi$ is a constant value 
at the mask. However, due to (b) the mask edge is a singular point in the sense that the contact
angle is not well defined there. In \cite{KBM2}, this situation was dealt with by instant readjustment
of the interface $-$ $x$-axis angle (angle of the tangent, $\phi_x$) 
to a new value once the contact point
reaches the mask edge to preserve the equilibrium constant angle $\phi$. Such instant readjustment causes the short
unphysical retraction of the contact point into the vapor phase and influences the shape of the interface 
\footnote{Angle of the tangent
is necessary for the determination of
the location of the contact point on the mask by coordinate's extrapolation from marker particles
adjacent to the contact point, through 
\begin{equation}
\tan{\phi_x} = \frac{dy/ds}{dx/ds},
\label{1.4}
\end{equation}
where $s$ is the arc length along the curve (crystal-vapor interface).
}.
In this paper we make an attempt to model
the overgrowth event self-consistently by approximating the mask edge by a circular arc 
and restricting the contact point to movement along this arc. We numerically study the limit of zero arc radius.
Since in applications the mask is often a crystalline solid itself, we allow for the anisotropy
of $\gamma_{mv}$ and thus, by virtue of \rf{1.3}, for the variable $\phi$ along the mask.  
(Quartz or tungsten masks are common \cite{MaukCu01}. 
Sometimes a mask is not deposited at all, but the long stripes are etched
in the crystalline GaAs or sapphire substrate itself \cite{THRUSH,Notzel,Hsu}. The long crystalline mesas
left on a surface of a substrate in this latter case are more advantageous for the epitaxy of some semiconductor 
materials than the 
geometrically equivalent mesas of the foreign (to the substrate) mask material.)

Apart from the interest to this problem 
that stems from the ELO technology, there is a broad class of phenomena in which the material-defect
interaction
is a key component. For instance, when liquid wets the microscopically rough surface
the contact line gets pinned on surface asperities and thus becomes unstable. The latter problem, both for a single 
asperity and for the random ensemble of asperities has been
studied intensively from a physico-mechanical standpoint;
see, for instance, recent works \cite{Quere,Golestanian} and
the review article by de Gennes \cite{DeGennes_review} 
. 
The somewhat different
situation where the surface defect is the macroscopic, organized structure 
was considered, for example, by Oliver \textit{et al.} in \cite{Oliver}. In that work, the defect is
the edge of a solid disk that supports 
the liquid drop on its top.
The pinning effect of the edge has been 
examined theoretically and Gibbs inequality condition for the equilibrium of a drop bound
by the edge has been confirmed experimentally. Also, the various conditions for drop stability
at the edge have been examined. Authors used a circular sapphire disk with the 90$^\circ$ edge and the
aluminum disks with the edges subtending a range of angles.
In \cite{LB1}, the finite element stability analysis of an inclined pendant drop
was performed, as well as the experiment. The latter showed that both the contact line and the contact
angle can adjust \textit{around} the capillary until the Young-Laplace equation 
that balances the capillary pressure
with pressure forces due to gravity \cite{DeGennes_review} is satisfied. 
In \cite{OBrien}, the asymptotic solution
to the model of the contact line pinning on a single, macroscopic spherical defect was derived.
However, our bibliography search for the study in which the direct numerical tracking of the contact line
motion over the defect is performed was unsuccessful.
Obviously, the ELO problem 
for the homoepitaxial crystal is simpler 
than any similar problem involving a liquid, since 
there is at least no need to solve the Young-Laplace equation. 
Heteroepitaxial and thermal stresses, if present in a
solid-on-solid system, may significantly complicate the analysis, but they
usually are incapable of modifying the equilibrium contact angle determined by the interfacial energies alone
\cite{SD} (however, as is well known, they may alter the dynamics of the free surface of a solid film).
The latter in the case of the SAG was demonstrated in \cite{EVpatt}, 
where the morphological evolution of a heteroepitaxial thin 
film growing on a patterned substrate was studied by the phase-field method.

\Section{Formulation}

\label{Sec2}

\begin{figure}
\setlength{\unitlength}{0.240900pt}
\ifx\plotpoint\undefined\newsavebox{\plotpoint}\fi
\sbox{\plotpoint}{\rule[-0.200pt]{0.400pt}{0.400pt}}%
\begin{picture}(1500,900)(0,0)
\font\gnuplot=cmr10 at 10pt
\gnuplot
\sbox{\plotpoint}{\rule[-0.200pt]{0.400pt}{0.400pt}}%
\put(60.0,40.0){\rule[-0.200pt]{332.201pt}{0.400pt}}
\put(1439.0,40.0){\rule[-0.200pt]{0.400pt}{197.538pt}}
\put(60.0,860.0){\rule[-0.200pt]{332.201pt}{0.400pt}}
\put(129,122){\makebox(0,0){$h_m$}}
\put(290,106){\makebox(0,0){MASK}}
\put(979,73){\makebox(0,0){STRIPE (SUBSTRATE)}}
\put(428,171){\makebox(0,0){$(x_*,y_*)$}}
\put(359,237){\makebox(0,0){$s = 0$}}
\put(462,227){\makebox(0,0){$\phi$}}
\put(749,368){\makebox(0,0){CRYSTAL}}
\put(1324,719){\makebox(0,0){$s = S(t)$}}
\put(520,7){\makebox(0,0){$(0,0)$}}
\put(60,7){\makebox(0,0){$-\ell$}}
\put(1439,7){\makebox(0,0){$L$}}
\put(1186,420){\makebox(0,0){$x$}}
\put(1071,516){\makebox(0,0){$y$}}
\put(313,614){\makebox(0,0){VAPOR}}
\put(1048,745){\makebox(0,0){${\mathbf q}_c$}}
\put(1094,719){\makebox(0,0){$\alpha$}}
\put(143,227){\makebox(0,0){$\theta$}}
\put(96,258){\makebox(0,0){${\mathbf q}_m$}}
\put(60.0,40.0){\rule[-0.200pt]{0.400pt}{197.538pt}}
\put(175,40){\vector(0,1){164}}
\put(175,204){\vector(0,-1){164}}
\put(1094,450){\vector(0,1){82}}
\put(1094,450){\vector(1,0){115}}
\put(1071,696){\vector(0,1){82}}
\put(129,204){\vector(0,1){82}}
\sbox{\plotpoint}{\rule[-0.500pt]{1.000pt}{1.000pt}}%
\put(438,204){\usebox{\plotpoint}}
\multiput(438,204)(2.836,20.561){2}{\usebox{\plotpoint}}
\put(443.67,245.12){\usebox{\plotpoint}}
\multiput(446,262)(2.836,20.561){2}{\usebox{\plotpoint}}
\put(452.90,306.69){\usebox{\plotpoint}}
\put(456.36,327.15){\usebox{\plotpoint}}
\multiput(459,345)(3.779,20.409){2}{\usebox{\plotpoint}}
\put(466.55,388.57){\usebox{\plotpoint}}
\put(470.20,408.99){\usebox{\plotpoint}}
\put(474.32,429.34){\usebox{\plotpoint}}
\multiput(478,447)(4.409,20.282){2}{\usebox{\plotpoint}}
\put(487.39,490.21){\usebox{\plotpoint}}
\put(491.97,510.46){\usebox{\plotpoint}}
\put(496.73,530.66){\usebox{\plotpoint}}
\put(501.70,550.81){\usebox{\plotpoint}}
\put(506.92,570.90){\usebox{\plotpoint}}
\put(513.22,590.67){\usebox{\plotpoint}}
\put(519.18,610.55){\usebox{\plotpoint}}
\put(526.05,630.13){\usebox{\plotpoint}}
\put(533.40,649.53){\usebox{\plotpoint}}
\put(542.16,668.32){\usebox{\plotpoint}}
\put(551.44,686.89){\usebox{\plotpoint}}
\put(562.71,704.28){\usebox{\plotpoint}}
\put(575.60,720.50){\usebox{\plotpoint}}
\put(592.42,732.61){\usebox{\plotpoint}}
\put(611.94,739.00){\usebox{\plotpoint}}
\put(632.64,738.23){\usebox{\plotpoint}}
\put(652.24,731.92){\usebox{\plotpoint}}
\put(671.27,723.87){\usebox{\plotpoint}}
\put(689.83,714.59){\usebox{\plotpoint}}
\put(708.58,705.71){\usebox{\plotpoint}}
\put(727.82,698.06){\usebox{\plotpoint}}
\put(747.80,692.64){\usebox{\plotpoint}}
\put(768.25,689.13){\usebox{\plotpoint}}
\put(788.99,689.00){\usebox{\plotpoint}}
\put(809.75,689.00){\usebox{\plotpoint}}
\put(830.34,691.00){\usebox{\plotpoint}}
\put(850.91,693.00){\usebox{\plotpoint}}
\put(871.50,695.00){\usebox{\plotpoint}}
\put(892.17,696.03){\usebox{\plotpoint}}
\put(912.85,697.00){\usebox{\plotpoint}}
\put(933.52,698.00){\usebox{\plotpoint}}
\put(954.23,697.46){\usebox{\plotpoint}}
\put(974.95,697.00){\usebox{\plotpoint}}
\put(995.71,697.00){\usebox{\plotpoint}}
\put(1016.38,696.00){\usebox{\plotpoint}}
\put(1037.13,696.00){\usebox{\plotpoint}}
\put(1057.89,696.00){\usebox{\plotpoint}}
\put(1078.65,696.00){\usebox{\plotpoint}}
\put(1099.40,696.00){\usebox{\plotpoint}}
\put(1120.16,696.00){\usebox{\plotpoint}}
\put(1140.91,696.00){\usebox{\plotpoint}}
\put(1161.67,696.00){\usebox{\plotpoint}}
\put(1182.42,696.00){\usebox{\plotpoint}}
\put(1203.18,696.00){\usebox{\plotpoint}}
\put(1223.93,696.00){\usebox{\plotpoint}}
\put(1244.69,696.00){\usebox{\plotpoint}}
\put(1265.44,696.00){\usebox{\plotpoint}}
\put(1286.20,696.00){\usebox{\plotpoint}}
\put(1306.96,696.00){\usebox{\plotpoint}}
\put(1327.71,696.00){\usebox{\plotpoint}}
\put(1348.47,696.00){\usebox{\plotpoint}}
\put(1369.22,696.00){\usebox{\plotpoint}}
\put(1389.98,696.00){\usebox{\plotpoint}}
\put(1410.73,696.00){\usebox{\plotpoint}}
\put(1431.49,696.00){\usebox{\plotpoint}}
\put(1439,696){\usebox{\plotpoint}}
\sbox{\plotpoint}{\rule[-0.200pt]{0.400pt}{0.400pt}}%
\multiput(60,204)(20.756,0.000){23}{\usebox{\plotpoint}}
\multiput(520,204)(0.000,-20.756){8}{\usebox{\plotpoint}}
\put(520,40){\usebox{\plotpoint}}
\put(520,40){\usebox{\plotpoint}}
\put(520,40){\usebox{\plotpoint}}
\put(520,40){\usebox{\plotpoint}}
\put(520.0,40.0){\rule[-0.200pt]{0.400pt}{4.818pt}}
\put(519.0,60.0){\usebox{\plotpoint}}
\put(518,90.67){\rule{0.241pt}{0.400pt}}
\multiput(518.50,90.17)(-0.500,1.000){2}{\rule{0.120pt}{0.400pt}}
\put(519.0,60.0){\rule[-0.200pt]{0.400pt}{7.468pt}}
\put(518,92){\usebox{\plotpoint}}
\put(518,92){\usebox{\plotpoint}}
\put(518,92){\usebox{\plotpoint}}
\put(518,92){\usebox{\plotpoint}}
\put(517,108.67){\rule{0.241pt}{0.400pt}}
\multiput(517.50,108.17)(-0.500,1.000){2}{\rule{0.120pt}{0.400pt}}
\put(518.0,92.0){\rule[-0.200pt]{0.400pt}{4.095pt}}
\put(517,110){\usebox{\plotpoint}}
\put(517,110){\usebox{\plotpoint}}
\put(517,110){\usebox{\plotpoint}}
\put(517,110){\usebox{\plotpoint}}
\put(517.0,110.0){\rule[-0.200pt]{0.400pt}{3.132pt}}
\put(516.0,123.0){\usebox{\plotpoint}}
\put(516.0,123.0){\rule[-0.200pt]{0.400pt}{2.650pt}}
\put(515.0,134.0){\usebox{\plotpoint}}
\put(514,142.67){\rule{0.241pt}{0.400pt}}
\multiput(514.50,142.17)(-0.500,1.000){2}{\rule{0.120pt}{0.400pt}}
\put(515.0,134.0){\rule[-0.200pt]{0.400pt}{2.168pt}}
\put(514,144){\usebox{\plotpoint}}
\put(514,144){\usebox{\plotpoint}}
\put(514,144){\usebox{\plotpoint}}
\put(514,144){\usebox{\plotpoint}}
\put(514,144){\usebox{\plotpoint}}
\put(514.0,144.0){\rule[-0.200pt]{0.400pt}{1.927pt}}
\put(513.0,152.0){\usebox{\plotpoint}}
\put(513.0,152.0){\rule[-0.200pt]{0.400pt}{1.686pt}}
\put(512.0,159.0){\usebox{\plotpoint}}
\put(512.0,159.0){\rule[-0.200pt]{0.400pt}{1.445pt}}
\put(511.0,165.0){\usebox{\plotpoint}}
\put(511.0,165.0){\rule[-0.200pt]{0.400pt}{1.445pt}}
\put(510.0,171.0){\usebox{\plotpoint}}
\put(510.0,171.0){\rule[-0.200pt]{0.400pt}{1.204pt}}
\put(509.0,176.0){\usebox{\plotpoint}}
\put(509.0,176.0){\rule[-0.200pt]{0.400pt}{1.204pt}}
\put(508.0,181.0){\usebox{\plotpoint}}
\put(508.0,181.0){\rule[-0.200pt]{0.400pt}{0.964pt}}
\put(507.0,185.0){\usebox{\plotpoint}}
\put(507.0,185.0){\rule[-0.200pt]{0.400pt}{0.723pt}}
\put(506.0,188.0){\usebox{\plotpoint}}
\put(505,190.67){\rule{0.241pt}{0.400pt}}
\multiput(505.50,190.17)(-0.500,1.000){2}{\rule{0.120pt}{0.400pt}}
\put(506.0,188.0){\rule[-0.200pt]{0.400pt}{0.723pt}}
\put(505,192){\usebox{\plotpoint}}
\put(505,192){\usebox{\plotpoint}}
\put(505,192){\usebox{\plotpoint}}
\put(505,192){\usebox{\plotpoint}}
\put(505,192){\usebox{\plotpoint}}
\put(505,192){\usebox{\plotpoint}}
\put(505,192){\usebox{\plotpoint}}
\put(505,192){\usebox{\plotpoint}}
\put(505,192){\usebox{\plotpoint}}
\put(505,192){\usebox{\plotpoint}}
\put(505,192){\usebox{\plotpoint}}
\put(505,192){\usebox{\plotpoint}}
\put(505.0,192.0){\rule[-0.200pt]{0.400pt}{0.482pt}}
\put(504.0,194.0){\usebox{\plotpoint}}
\put(504.0,194.0){\rule[-0.200pt]{0.400pt}{0.723pt}}
\put(503.0,197.0){\usebox{\plotpoint}}
\put(503.0,197.0){\rule[-0.200pt]{0.400pt}{0.482pt}}
\put(502.0,199.0){\usebox{\plotpoint}}
\put(502.0,199.0){\usebox{\plotpoint}}
\put(501.0,200.0){\usebox{\plotpoint}}
\put(501.0,200.0){\rule[-0.200pt]{0.400pt}{0.482pt}}
\put(500.0,202.0){\usebox{\plotpoint}}
\put(500.0,202.0){\usebox{\plotpoint}}
\put(498.0,203.0){\rule[-0.200pt]{0.482pt}{0.400pt}}
\put(498.0,203.0){\usebox{\plotpoint}}
\put(60.0,204.0){\rule[-0.200pt]{105.514pt}{0.400pt}}
\end{picture}
\vspace*{0.7cm}
\caption{A sketch of the mathematical situation for the crystal growth
on the stripe-patterned, masked substrate.  
The crystal-vapor interface is defined parametrically as 
$y=y(s,t),\ x=x(s,t),\ 0 \le s \le S(t)$, where $s$ is the arc length
along the curve and $S$ is the total arc length of the curve.
Due to periodical
arrangement of the stripes, the crystal growth is studied on a partial cross-section which is a line
segment extending from the center line of one mask
surface at $x=-\ell$ to the center of the adjacent stripe at $x=L$.
The interface
is sketched such that the crystal overgrowth on the mask
is 
shown. The mask is assumed rectangular, with the rounded edge (solid curve; since the lateral length scale
in the picture is much larger than the vertical one, the edge in the form of the circular arc does not 
look circular).
$\alpha$ is the angle that the outward unit normal to the interface, ${\mathbf q}_c$, makes \textit{with 
the
horizontal axis}. 
$\phi$ is the contact angle that the interface makes \textit{with the mask surface}, measured from within
the crystal. $\theta$ is the angle that the unit normal to the mask surface, ${\mathbf q}_m$, makes \textit{with 
the
horizontal axis}. $(x_*,y_*)$ is the contact point.
} 
\label{Fig1}
\end{figure}
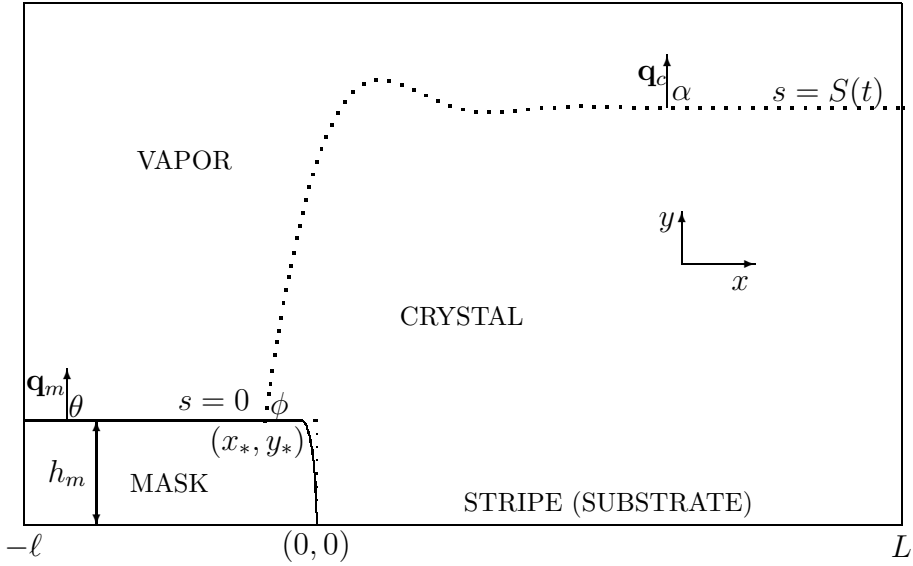

The Figure \ref{Fig1} shows the problem geometry. 
As was already noted, the length of the open, unmasked stripes is assumed much larger 
than their widths since large aspect ratios are common
in the applications; thus, the 2D crystal growth model can be used.
%

\subsection{Equations of crystal growth}

Based on the classical Mullins' approach \cite{MULLINS57}, the normal velocity of the
interface is formulated as a function of the local curvature and its derivatives.  
In this work, we use 
\begin{equation}
V_q = D \frac{\partial^2}{\partial s^2}\left[\left(\hat \gamma_{cv}(\alpha)+
\frac{\partial^2 \hat \gamma_{cv}(\alpha)}{\partial \alpha^2}\right) 
K\right] + J\hat M(\alpha),
\label{1.0}
\end{equation}
which is the simplified expression from \cite{KBM2}. We neglect the evaporation-recondensation term
as being unimportant for this study and assume the isotropic interfacial diffusivity. 
The equation \rf{1.0} is in the non-dimensional form.
$D$ has the meaning of the constant effective interfacial diffusivity (see Table \ref{Table2} in the Appendix B), 
$K$ is the curvature and $J$ the composite parameter that involves, in particular,
the chemical potential of the vapor phase and the mean value of the interface mobility, $M_0$ (the 
kinetic coefficient).
$\hat \gamma_{cv}(\alpha)$ and $\hat M(\alpha)$ are the anisotropy factors of the 
interface energy and mobility, respectively. 
These factors are specified in Section \ref{azotr}.
The
marker particles on the interface are numerically advanced using the parametric evolution equations
for the Cartesian coordinates of particles, viz.
\begin{equation}
\label{1.1}
\frac{\partial x}{\partial t} = V_q\ \frac{\partial y}{\partial s},
\end{equation}
$$
\frac{\partial y}{\partial t} = -V_q\ \frac{\partial x}{\partial s}.
$$
These equations are fourth-order parabolic when $V_q$ is given by \rf{1.0}.

\subsection{Mask Energy and Shape}

\begin{figure}[H]
\centering
\psfig{figure=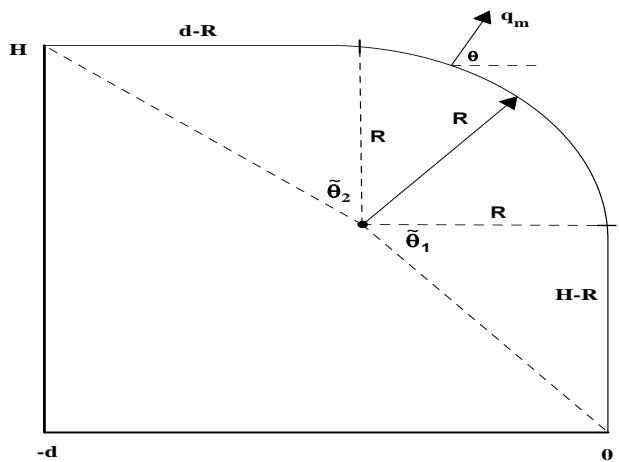,height=3.6in,width=4.0in,angle=0}
\caption{Half of the mask with the rounded edge.}
\label{Fig2}
\end{figure}

\begin{figure}[H]
\centering
\psfig{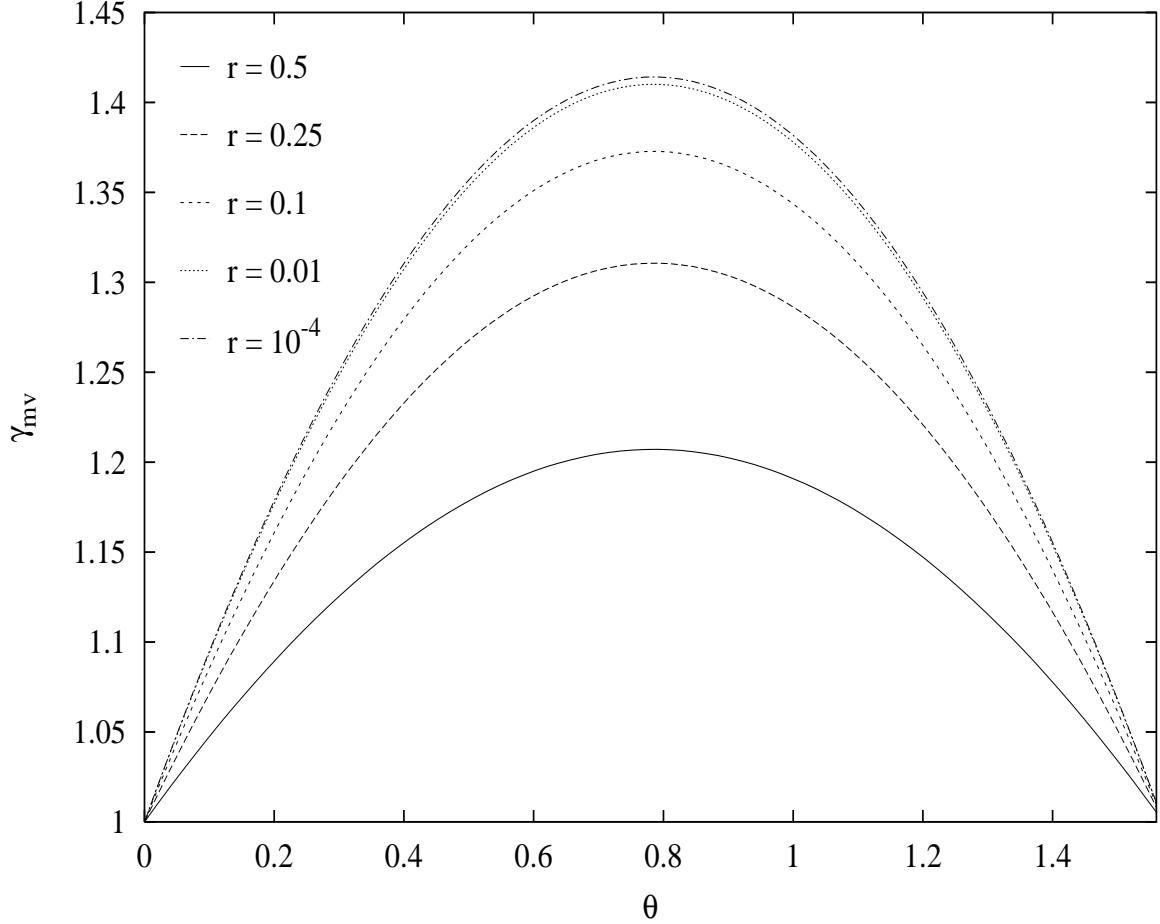}
\caption{The nondimensional energy of the mask surface as function of the angle of the unit 
normal (to that surface), for the different
ratios of the radius of the circular arc to the mask height. The points $(0,1)$ and $(\pi/2,1)$ correspond to
the vertical and horizontal parts of the mask, respectively; other points correspond to the circular arc
that connects the vertical and horizontal parts.}
\label{gammaplot}
\end{figure}

To derive the expression for $\gamma_{mv}$ which accounts for the additional edge energy, 
we use the method of
Xin \& Wong \cite{XinWong01,XinWongreport}, which they employed in \cite{XinWongGB} for studies of the grain boundary 
(GB) grooving in polycrystalline
thin films. In contrast to the case of dynamically evolving, faceted
interfaces such as GB grooves, the application to the static crystalline mask surface is straightforward 
since the method requires an \textit{a priori}\ knowledge of facets orientations and lengths.
Details of the method can be found in \cite{XinWong01,XinWongreport}. 
For the rectangular crystal in the thermodynamic equilibrium that represents the mask on a substrate
(half of that crystal is shown in Fig. \ref{Fig1},\ref{Fig2}), the result is the following nondimensional 
expression (units for the surface 
energy are chosen $\mu_m h_m/\Omega$, where $\Omega$ is the atomic volume and $\mu_m$ is the equilibrium 
chemical potential
of the mask surface):
\begin{equation}
\gamma_{mv} = r + \left[\bar \ell - r + 2(r-\bar \ell)F\left(\theta - \frac{\pi}{2}\right)\right]\cos{\theta} +
2(1-r)\left[\frac{1}{2} - F\left(\theta - \pi\right)\right]\sin{\theta},\quad 0 \le \theta \le \pi.
\label{2.1}
\end{equation} 
In \rf{2.1}, $r=R/H$ is the ratio of the nondimensional circular arc radius to the nondimensional 
step height, $\bar \ell = d/H$, where $d$ is the nondimensional half-width of the mask, and
$F(u)$ is Heaviside function (all lengths are nondimensionalized by the half-width of the stripe, $L$). 
For our modeling of the ELO, we are only interested in values of
$\theta$ in $[0,\pi/2]$. The function $\gamma_{mv}(\theta)$ is plotted in Fig. \ref{gammaplot} for $\bar \ell = 1$
and decreasing values
of $r$. Notice that $\gamma_{mv}$ is at maximum at $\theta=\pi/4$, 
since the half-mask is taken square with the rounded edge. 
The sharper is the edge, the larger is the maximum value. For the 90$^\circ$
edge $r$ is exactly zero, $\theta$ takes on values 0 and $\pi/2$ only, and \rf{2.1} 
gives $\gamma_{mv}(0)=\gamma_{mv}(\pi/2)=1$.


To parametrize the mask surface, we take angle $\tilde \theta \in 
\left[-\tilde \theta_1, \tilde \theta_2 + \pi/2\right]$ 
as the independent 
variable (see Fig. \ref{Fig2}). From geometry,
\begin{equation}
\tilde \theta_1 = \mbox{arctan} \frac{H-R}{R} = \mbox{arctan}\left(\frac{1}{r}-1\right),
\label{2.2}
\end{equation}
$$
\tilde \theta_2 = \mbox{arctan} \frac{d-R}{R} = \mbox{arctan}\left(\frac{d}{R}-1\right).
$$
Then,
\begin{equation}
\theta = \left\{
\begin{array}{l}
0, \mbox{ if}\;\; -\tilde \theta_1\le \tilde \theta < 0,  \label{2.3} \\
\tilde \theta, \mbox{ if}\;\; 0 \le \tilde \theta \le \pi/2, \\
\pi/2, \mbox{ if}\;\; \pi/2 < \tilde \theta \le \pi/2 + \tilde \theta_2. 
\end{array}
\right. 
\end{equation}
Also, 
\begin{equation}
x_m = \left\{
\begin{array}{l}
0, \mbox{ if}\;\; -\tilde \theta_1\le \tilde \theta < 0,  \label{2.4} \\
R\left(-1+\cos{\tilde \theta}\right), \mbox{ if}\;\; 0 \le \tilde \theta \le \pi/2, \\
-R\left(1+\tan{\tilde \theta}\right), \mbox{ if}\;\; \pi/2 < \tilde \theta \le \pi/2 + \tilde \theta_2, 
\end{array}
\right. 
\end{equation}
\begin{equation}
y_m = \left\{
\begin{array}{l}
H - R\left(1+\tan{\tilde \theta}\right), \mbox{ if}\;\; -\tilde \theta_1\le \tilde \theta < 0,  \label{2.5} \\
H - R\left(1-\sin{\tilde \theta}\right), \mbox{ if}\;\; 0 \le \tilde \theta \le \pi/2, \\
H, \mbox{ if}\;\; \pi/2 < \tilde \theta \le \pi/2 + \tilde \theta_2 
\end{array}
\right. 
\end{equation}
are the nondimensional parametric functions that describe the mask surface. From \rf{2.3}-\rf{2.5}, 
the rounded portion of the mask is described by
\begin{equation}
\left\{
\begin{array}{l}
x_m^{(arc)} = R\left(-1+\cos{\theta}\right),  \label{2.6} \\
y_m^{(arc)} = H - R\left(1-\sin{\theta}\right), \; 0 \le \theta = \tilde \theta \le \pi/2.
\end{array}
\right. 
\end{equation}

Taking $\mu_m h_m/\Omega$ for the unit of $\gamma_{cm}$, and $\gamma_0$ (the mean value, see \rf{2.12} below) for 
the unit
of $\gamma_{cv}$,  the equation \rf{1.3} takes the form
\begin{equation}
E(\gamma_{mv}-\gamma_{cm}) - \hat \gamma_{cv}(\alpha^*)\cos{\phi} + 
\frac{\partial \hat \gamma_{cv}}{\partial \phi}(\alpha^*)\sin{\phi} = 0,
\label{2.1a}
\end{equation}
where $E=\mu_m h_m/(\Omega \gamma_0)$, $\gamma_{mv}$ is given by \rf{2.1},
$\gamma_{cm}$ now is the nondimensional energy of the crystal-mask interface (assumed constant and isotropic), 
$\hat \gamma_{cv}$ is the anisotropy
factor of the interfacial energy, and $\alpha^*$ is value of the angle of the normal to the interface at the 
tri-junction. Notice that
\begin{equation}
\alpha^* = \phi+\theta.
\label{2.1b}
\end{equation}
\rf{2.1a}
is solved once in the beginning of the computation and yields values of the \textit{imposed} contact angle at every
$\tilde \theta_i \in \left[-\tilde \theta_1, \tilde \theta_2 + \pi/2\right], i=1...N_{\tilde \theta}$,
where $N_{\tilde \theta}$ is the number of equidistant nodes in the interval.
$N_{\tilde \theta}$ is taken in the range 500...3000, the larger the smaller is the radius of the circular arc
(since for the accurate marching of the contact point along the arc it is important to have sufficient
number of nodes there).

\subsection{Updating the contact point}

In this section, the procedure which updates the position of the contact point on the mask is described. 
The update is needed
after all marker particles on the interface but the contact point are moved one time step forward
using the evolution equations \rf{1.1}.

Let $x_*$ and $y_*$ are the ``old" coordinates of the contact point. Using $x_*, y_*$, the coordinates of the
two nearest nodes on the mask: $x_m(i), y_m(i), x_m(i+1), y_m(i+1)$, and $\theta_{i,i+1},\phi_{i,i+1}$,
we first compute values of $\theta$ and $\phi$ at $(x_*,y_*)$ by linear interpolation. Using the latter values,
the angle of the tangent to the interface at $(x_*,y_*)$ is found from
\begin{equation}
\phi_x = \phi+\theta - \pi/2.
\label{2.7}
\end{equation}
Let $(x_1,y_1)$ and $(x_2,y_2)$ are the coordinates of the first two marker particles on the interface
closest to the contact point.
In case the contact point is on the vertical part of the mask, the updated location is $(0,\hat y_*)$, where
\begin{equation}
\hat y_* = \frac{1}{3}\left(4y_1-y_2 + \left(x_2-4x_1\right)\tan{\phi_x}\right).
\label{2.8}
\end{equation}
This follows from the second-order, one-sided finite difference approximation of the contact angle condition \rf{1.4}.
In case the contact point is on the horizontal part of the mask, the updated location is $(\hat x_*,H)$, where
\begin{equation}
\hat x_* = \frac{1}{3}\left(4x_1-x_2 + \frac{3H-4y_1+y_2}{\tan{\phi_x}}\right).
\label{2.9}
\end{equation}
In case the contact point is on the circular arc, we 
solve the quadratic equation (ref. \rf{2.6})
\begin{equation}
\frac{1}{3}\left(4y_1-y_2 + \left(3x_*-4x_1+x_2\right)\tan{\phi_x}\right) = 
H - R\left(1-\sqrt{1-(1+x_*/R)^2}\right)
\label{2.10}
\end{equation}
and find two $x$-coordinates  $\hat x_*^{(1)},\hat x_*^{(2)}$ and corresponding $y$-coordinates  
$\hat y_*^{(1)},\hat y_*^{(2)}$. Then, if both $\hat x_*^{(1)},\hat x_*^{(2)} \ge -R$ (that is, if
both possible updated locations are on the arc), we take the pair of coordinates closest to $(x_*,y_*)$
for the updated location.
If only one of $(\hat x_*^{(1},\hat y_*^{(1)}),\;(\hat x_*^{(2)},\hat y_*^{(2)})$ is on the arc and the other 
is not, we choose
the pair that is on the arc (closer to the ``old" location).

\subsection{Anisotropy}
 
\label{azotr}
The anisotropy factor of the kinetic mobility is chosen as in \cite{US}, where the following form is
suggested and the simulations of faceted growth are performed within the framework of phase field bulk 
solidification model:
\begin{equation}
\hat M(\alpha) = 1 - \sigma + 2\sigma\tanh{\frac{\chi}{|\tan{2\left(\alpha+\beta_M\right)}|}};\quad 
M = M_0 \hat M(\alpha).
\label{2.11}
\end{equation}
In \rf{2.11}, $\sigma$ and $\chi$ are constant parameters that control the width and flatness of
facets, respectively; $\beta_M$ is the phase angle. Facets are formed at regions on the interface
where $\alpha = \pi/4 - \beta_M + n\pi/2,\; n=0,1,...$. It was shown in \cite{US} that the results of computations
with the four-fold anisotropy \rf{2.11} are in good agreement with the so-called kinetic Wulff shapes 
(see references in \cite{US}; specifically for the selective area epitaxy the kinetic Wulff shapes
were constructed and compared to the experimental shapes in \cite{Jones}). For the simulations discussed
in the next section we chose $\sigma = 0.95,\; \chi = 2,\; \beta_M = 0$.

The anisotropy factor of the crystal-vapor interfacial energy has standard, four-fold symmetric form
\begin{equation}
\hat \gamma_{cv}(\alpha) = 1+\epsilon_\gamma \cos{ \left[4\left(\alpha+\beta_\gamma\right)\right] };\quad \gamma_{cv} =
\gamma_0 \hat \gamma_{cv}(\alpha),
\label{2.12}
\end{equation}
where the constant $\epsilon_\gamma > 0$ determines the degree of anisotropy, and $\beta_\gamma$ is the phase angle.
In this work we allow for strong anisotropy of $\gamma_{cv}$, e.g. $\epsilon_\gamma > 1/15$.
In this case, the interface stiffness $G =\hat \gamma_{cv}+\partial^2 \hat \gamma_{cv}/\partial \alpha^2$ is 
negative in 
certain intervals of values of $\alpha$, and that manifests in the appearance of
corners on the equilibrium Wulff shape of a crystal \cite{HERRING}; 
in the dynamical case, the evolution equations become backward-parabolic and unstable 
where $G<0$, see for example \cite{CGP} - \cite{SGDNV}.

\subsubsection{Regularization}

To track interface evolution with the strongly anisotropic interfacial energy, the problem must be regularized
in order to penalize spatial oscillations as well as the tendency to form corners. The regularization
we use in this work is based on the addition of curvature dependence to $\gamma_{cv}$. Regularization by curvature 
was first proposed
by Herring \cite{HERRING} for the equilibrium shape; see \cite{Brian_regular} for the modern mathematical analysis
of this
problem. Apart from being 
useful as a modeling tool, the regularization by curvature has physical ground, namely the interaction 
of steps on a crystal surface in the vicinity of a corner \cite{GDN}.
Unlike \cite{GDN,SGDNV} and other works where the $\delta \kappa^2, \delta = const.$ ($\kappa$ is the
dimensional curvature) term is added to 
the surface energy on the problem formulation stage,
here we regard the regularization only as a numerical technique intended to keep the computation alive, and 
regularize by adding same
term to $\gamma_{cv}$ given by \rf{2.12} only at the distinct moments and only at the locations on the 
interface where the 
instability (tendency to form a corner) occurs. This is done as follows. 

First, we continuously check the 
sign of $G$ at every marker location on the interface. Second, once at the certain location $G$ becomes negative
(the corner has been formed), we interrupt the computation, restore all variables including the local interface 
shape
and curvature to their previous states at positive $G$, add $\Delta K^2,\ \Delta = \delta/(\gamma_0 L^2)$ term 
to the nondimensional interfacial energy 
$\hat \gamma_{cv}(\alpha)$
at this location and 
solve the 
linear equation $G(\Delta)=0$. The result,
\begin{equation}
\Delta = \frac{15 \epsilon_\gamma\cos{ \left[4\left(\alpha+\beta_\gamma\right)\right] } - 1}{K^2 + 
2K\partial^2 K/\partial \alpha^2 + 2(\partial K/\partial \alpha)^2},
\label{2.12a}
\end{equation} 
is deflected by 0.1$\Delta$ to ensure  positive $G$.
The computation is then restarted with the regularized
$\gamma_{cv}$; if needed, the procedure is repeated. This method allows to reduce the influence of the regularization 
on interface shape to a minimum by (i) applying the regularization selectively and (ii) choosing value for the
regularization constant $\Delta$ no larger than needed to allow the computation to proceed. 
The maximum value $\Delta \sim 10^{-5}$ comes out from the computation, which translates into $\delta \sim 10^{-11}$
erg; thus the additional corner energy necessary to regularize the problem is indeed very small.
\footnote{In \cite{GDN}, parameter $\delta$ is introduced as the second partial derivative of the crystalline 
step energy
with respect to the first coordinate derivative of step density; this is difficult, if not impossible to extract
from experimental data. We are not aware of a single experiment there this or related quantity is measured.
Since value is unknown, we think it makes sense to find the suitable numerical 
approximation to the nondimensional counterpart as explained. 
}

It must be noted here that the numerical method itself introduces some minor smoothing, since the interface is remeshed
every time step with the help of cubic splines \cite{KBM1}.

The regularization by curvature is not used at the contact point and its vicinity. Instead, 
to ensure stable tracking of the sensitive contact point motion over the mask 
edge in the simulations with strongly anisotropic surface energy, we had to resort to the method which is
described in the next subsection.

\subsubsection{Contact point regularization}

We take 
\begin{equation}
\epsilon_\gamma = \epsilon_\gamma^{(0)} + \left(\Upsilon - \epsilon_\gamma^{(0)}\right)H_\lambda(s-s_c),
\label{2.13} 
\end{equation}
where $\epsilon_\gamma^{(0)}$ is constant value $\epsilon_\gamma$ takes on the interval of the arc 
length $0 \le s < s_c - \lambda$, 
$\Upsilon > \epsilon_\gamma^{(0)}$ is constant value $\epsilon_\gamma$ takes on the 
interval $s_c + \lambda < s \le S$, and
$H_\lambda(s-s_c)$ is the smoothed Heaviside function (see, for instance, \cite{SSO}), 
\textit{viz.}
\begin{equation}
H_\lambda (s-s_c) = \left\{
\begin{array}{l}
0, \mbox{ if}\; s - s_c < - \lambda,  \label{2.14} \\
\frac{1}{2}\left(1+\frac{s-s_c}{\lambda}+\frac{1}{\pi}\sin{\frac{\pi(s-s_c)}{\lambda}}\right), 
\mbox{ if}\; -\lambda \le s-s_c \le \lambda, \\
1, \mbox{ if}\; s - s_c > \lambda. 
\end{array}
\right. 
\end{equation}
In \rf{2.14}, $\lambda = const.$ is the small half-width of the transition region from $0$ to $1$ around
$s_c =const. \in
(0,S)$. \rf{2.13} and \rf{2.14}
imply that the intervals of the interface $0 \le s < s_c - \lambda$ and $s_c-\lambda \le s \le s_c + \lambda$ evolve 
with lesser degree of anisotropy
than the remaining interval $s_c + \lambda < s \le S$. For instance, for the simulation shown in
Fig. \ref{Fig8}, we chose  $\epsilon_\gamma^{(0)} = 0.01, s_c = 0.08, \lambda = 0.02$.
Of course,
such choice is difficult to justify physically, but nevertheless it allows to perform simulations and 
obtain the right interface shape some (small) distance away from the contact point \cite{Rich}.

\Section{Results}

\label{Sec4}

Since the parameter space is very large, the overgrowth was computed
for the separate cases of evolution with the anisotropic mobility and crystal-vapor interfacial energy.
We studied the influence of  parameters $r$ and $\gamma_{cm}$. Other parameters are fixed
at their values cited in Table \ref{Table2}.
 
\subsection{Overgrowth with anisotropic mobility}

It is assumed in this section that $\epsilon_\gamma = 0$, thus $\hat \gamma_{cv} \equiv 1$ and
the contact angle $\phi$ is determined from the equation (\ref{2.1a}) where the third term identically
equals zero.

\begin{figure}[H]
\centering
\psfig{figure=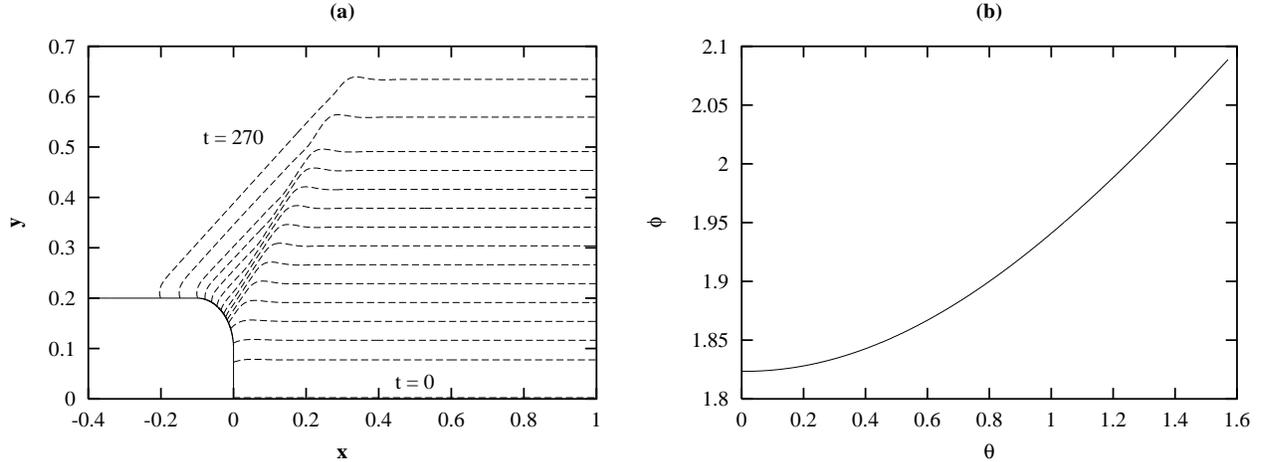,height=2.4in,width=6.5in,angle=270}
\caption{(a) The overgrowth over the blunt edge $r=0.5$ of the energetically 
inhomogeneous mask. Dashed curves show the interface at progressively increasing times. 
(b) The contact angle dependence on the angle of the normal to the mask surface, for
the crystal growth shown in (a).}
\label{Fig3}
\end{figure}

Fig. \ref{Fig3}(a) shows the overgrowth over the blunt edge $r=0.5$ (the radius = 50 nm) of the energetically 
inhomogeneous mask. 
Also shown in Fig. \ref{Fig3}(b)
is the \textit{actual} contact angle $\phi$ \textit{vs.}\ $\theta$. Value for $\gamma_{cm}$ is chosen 100 (500 erg/cm$^2$), so that $\phi(\theta) > \pi/2\  \forall\ \theta$ (partial ``wetting"). 
One of accuracy checks is the comparison every
time step of the actual value of the contact angle with the imposed $\phi(\theta)$
given by \rf{2.1a}. The difference is negligible for all runs performed.
Notice that the perfectly straight facet is formed in the direction 135$^\circ$ to the substrate.
The feature (``bump") is present at the junction of this facet and the horizontal, 90$^\circ$ facet.
This bump is present on the surface even before the contact point reaches the mask edge, but the edge
makes it more pronounced.  The bump gradually disappears after the edge is passed.

\begin{figure}[H]
\centering
\psfig{figure=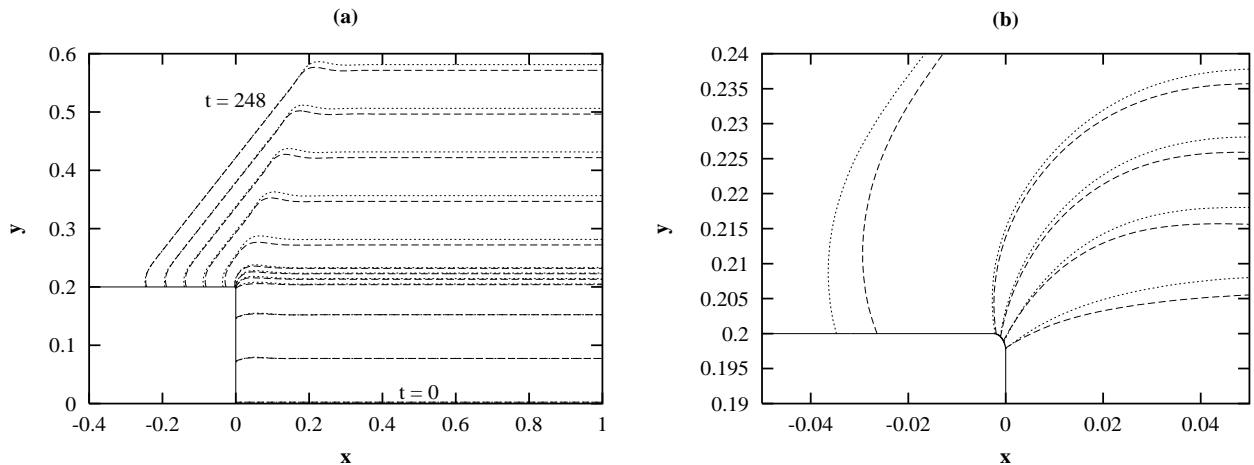,height=2.4in,width=6.5in,angle=270}
\caption{(a) The overgrowth over the sharp edge $r=0.0125$; dashed curve: $\phi(\theta)$ as in Fig. \ref{Fig3}(b), 
dotted curve: $\phi = 1.956$. (b) The enlarged view of the tri-junction region.}
\label{Fig4}
\end{figure}

Fig. \ref{Fig4}(a,b) compares the overgrowth over the sharper edge $r=0.0125$ (the radius = 1.25 nm) 
of the energetically inhomogeneous
mask as in Fig. \ref{Fig3}(b) with the overgrowth over
the edge of same sharpness of the energetically homogeneous mask. For the latter,
value for the constant $\phi$ is taken the average of the 
smallest and largest values of $\phi$ in  Fig. \ref{Fig3}(b) ($\phi = 1.956$). The overgrowth over the
energetically homogeneous mask is faster, and the bump is more pronounced in this case. 
Nevertheless, the difference in the overgrowth of the mask edges of different energy is quite small.

\begin{figure}[H]
\centering
\psfig{figure=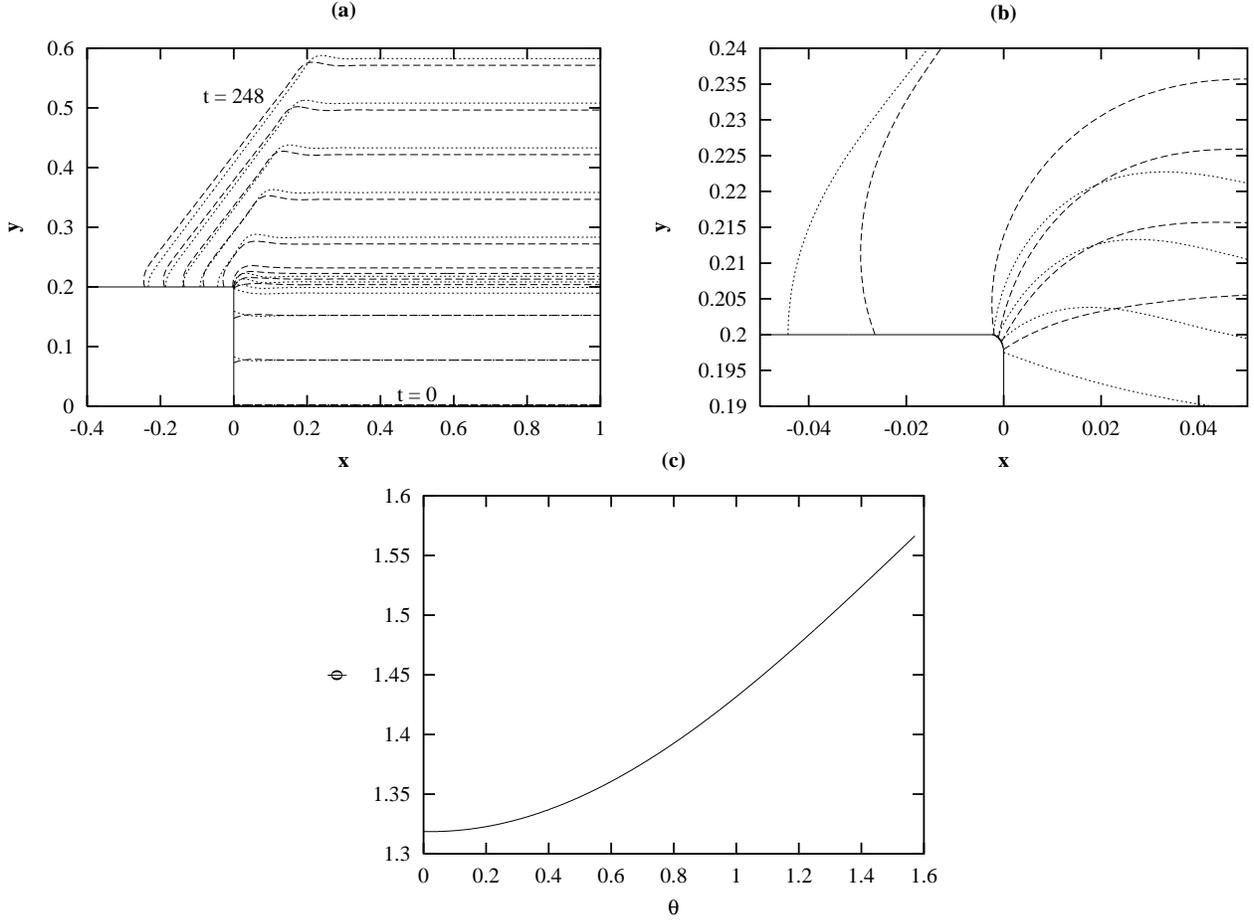,height=4.8in,width=6.5in,angle=270}
\caption{(a) The overgrowth over the sharp edge $r=0.0125$; dashed curve: $\phi(\theta)$ as in 
Fig. \ref{Fig3}(b) (less wetting), 
dotted curve: $\phi(\theta)$ as in part (c) of this figure (more wetting). 
(b) The enlarged view of the tri-junction region. (c) The contact angle \textit{vs.} $\theta$. To obtain this
dependence, value for $\gamma_{cm}$ is chosen 0.1 (0.5 erg/cm$^2$).}
\label{Fig5}
\end{figure}

Fig. \ref{Fig5}(a,b) compares the overgrowth over the $r=0.0125$ edge 
of the energetically inhomogeneous mask that supports different $\phi(\theta)$
(that is, the dynamics corresponding to $\phi(\theta) > \pi/2$ as in Fig. \ref{Fig3}(b), and 
$\phi(\theta) \le \pi/2$ as in Fig. \ref{Fig5}(c)). 
Interestingly,
the bump and the crystal thickness are larger for the less wetting case (larger $\phi$'s) 
when the contact point slides over the mask edge. However, immediately after the edge 
the overgrowth of the more wetting crystal is faster, but it slows down as the overgrowth
on the mask progresses. This is expected since $\phi = \pi/2$ on the horizontal part of the mask. 
Finally, the less wetting crystal takes over, but its 
thickness in the $y$-direction is less than one of the more wetting crystal.

\begin{figure}[H]
\centering
\psfig{figure=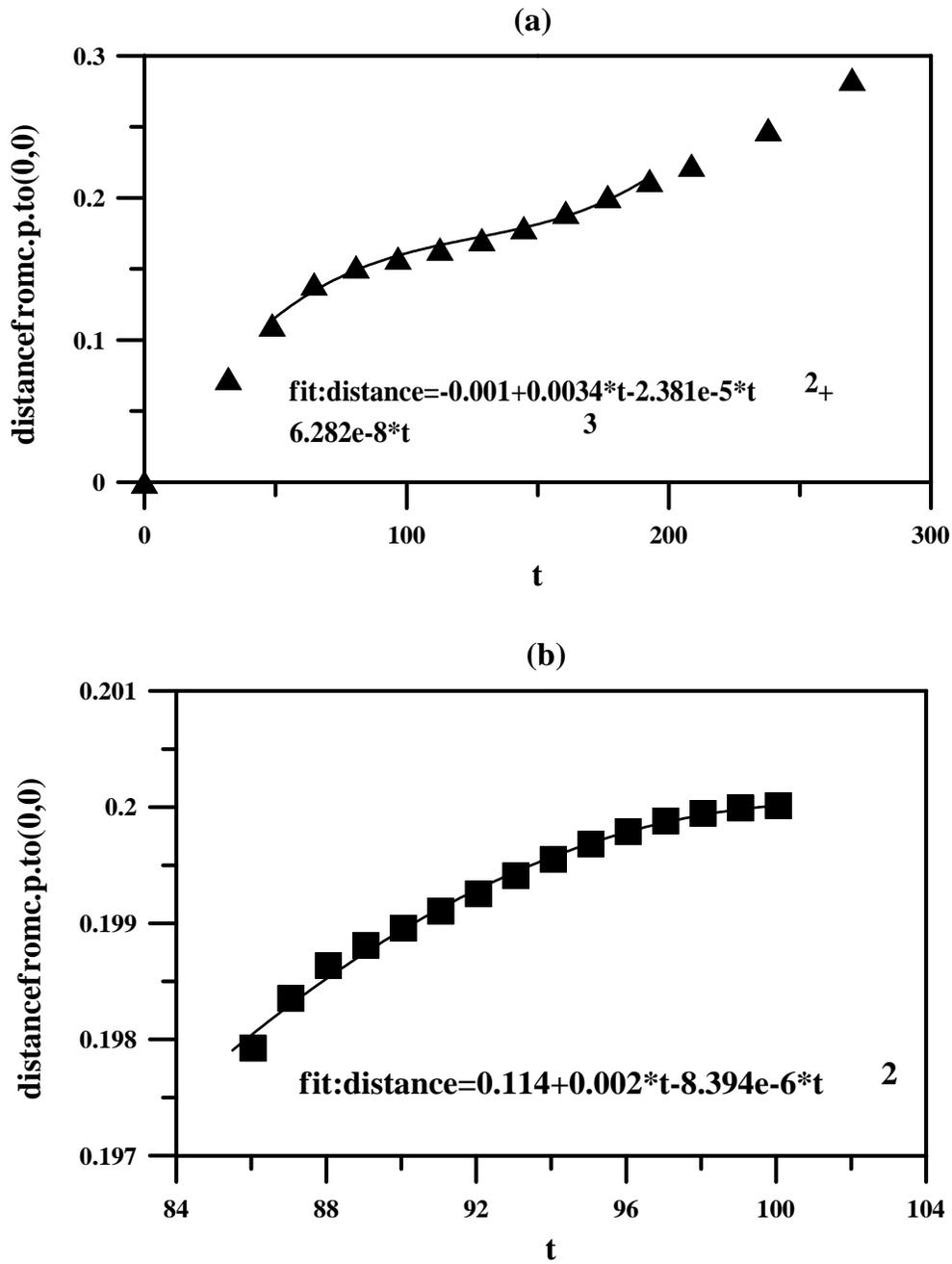,height=7.5in,width=5.8in,angle=0}
\caption{Distance \textit{vs.} time (nondimensional) from the tri-junction to the 
foot point (0,0) of the mask
for two values of the radius of curvature of the edge. 
(a) $r=0.5$ (triangular symbols). (b) $r=0.0125$ (square symbols). Fitting (solid curve)
is for the time interval when the contact point is on the circular arc of the mask.}
\label{Fig6}
\end{figure}

The distance from the contact point to the
foot point (0,0) of the mask \textit{vs.}\  time is shown in Fig. \ref{Fig6}(a,b) for 
the overgrowth
over $r=0.5$ and $r=0.0125$ edge, respectively, and $\gamma_{cm}=100$. 
For the blunt edge, the computational data is fitted well
by the cubic polynomial and thus the speed of the contact point \textit{vs.} time 
is the quadratic function that has
the minimum value $\approx 4\times 10^{-4}$ (4 nm/s) at $t=125$ (6.3 s) and the maximum value
$\approx 1.4\times 10^{-3}$ (14 nm/s) at $t=50,200$ (2.5,10 s). For the sharp edge,
the quadratic fit is better, and the speed of the contact point is  linear function of time
(the min. value $\approx 3.25\times 10^{-4}$ (3.25 nm/s) at $t=100$ (5 s),
the max. value $\approx 5.5\times 10^{-4}$ (5.5 nm/s) at $t=86$ (4.3 s)). 


\noindent
This data,
as well as Figures \ref{Fig3}-\ref{Fig5} demonstrate
that de-pinning of the contact line of a crystal from the solid defect of finite size does 
not have a threshold nature; 
that is, the contact line is in constant motion over the defect, and the problem does not have a critical 
eigenvalue such that the contact line is pinned when some parameter is smaller than this eigenvalue and 
de-pinned (moving) when the parameter is larger. 
In the language of \cite{DeGennes_review}, where the 
analogy between perturbed (on a defect) liquid contact line and an elastic spring is used to phenomenologically 
explain pinning (pages 836-38),
the total pinning force exerted by the defect on the line is such that graphs of this force and the restoring
linear elastic force (\textit{vs.} position on a defect) are not intersecting, and thus there is no 
equilibrium, pinning line position(s) on a defect.

\begin{figure}[H]
\centering
\psfig{figure=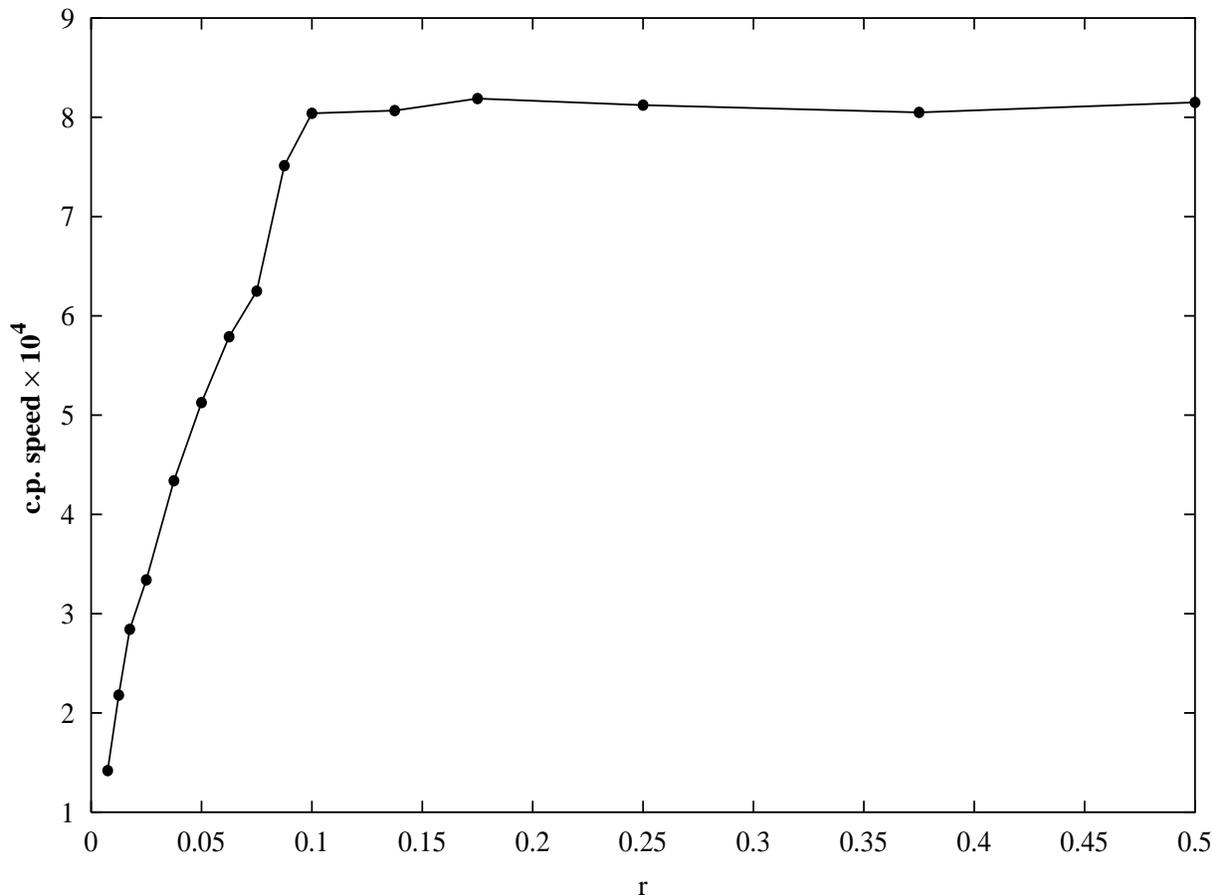,height=4.8in,width=6.5in,angle=270}
\caption{The average speed of the contact point's motion over the mask edge.}
\label{Fig7}
\end{figure}

Finally, the average nondimensional speed of the contact point's motion over the edge 
(naturally defined as the length of the circular segment of the mask, $\pi R/2$, 
divided by the traversal time) is plotted in Fig. \ref{Fig7} \textit{vs.}\ $r$.
There is an evident tendency for pinning of the contact point at the edge as $r \rightarrow 0$. 
Data can be satisfactory fitted by a hyperbolic tangent curve (not shown).
In the limit $r=0$ (the mathematically sharp edge)
the true pinning  is expected (the speed $=$ 0). The contact line starts to actually
``feel" the edge only when the latter has sufficiently small radius of curvature 
($r\approx 0.1-0.15\; \Leftrightarrow\;$10-15 nm). 
 
Note that despite such characteristic of the motion of the contact line over the edge as the speed
(instantaneous and average) is, of course, a function of all parameters of the system, the
pinning on the edge as its radius of curvature approaches zero is the universal tendency.
This was checked by means of multiple runs with different parameters. The computation fails
for $r < 0.0075$ (the radius$ < 0.75$ nm) because as the number of discretizing nodes along the mask increases,
the nodes on the two straight segments of the mask come too close to each other and the
interpolation becomes inaccurate. In principle, the computation with smaller radii can be made successful
by using the nonuniform grid along the mask; we did not attempt to do that. It is also doubtful that
very sharp mask edges (having the radii of curvature of the order of a molecule size (1-2 nm)) 
are achievable by the lithography, etching and polishing used in the ELO/SAG practice 
for the preparation of a mask. For comparison, the sapphire disks used for the late 1970s experiments with
the liquid drops by Oliver \textit{et al.} \cite{Oliver} had the edge radius of curvature smaller than 50 nm, 
and ``much smaller" than that
value if the defects (chipped-off hollows) appear along the edge. 
These disks were prepared by cutting from a fused sapphire boule. 
The aluminum disks they used had very blunted edges with the radius of curvature 1 to 5 $\mu$m.

\subsection{Overgrowth with strongly anisotropic $\gamma_{cv}$}

It is assumed in this section that $\sigma = 0$, thus $\hat M \equiv 1$. 
The contact angle $\phi$ is determined from the equation (\ref{2.1a}), where the third term is nonzero
due to the anisotropy in the crystal-vapor surface energy.

\begin{figure}[H]
\centering
\psfig{figure=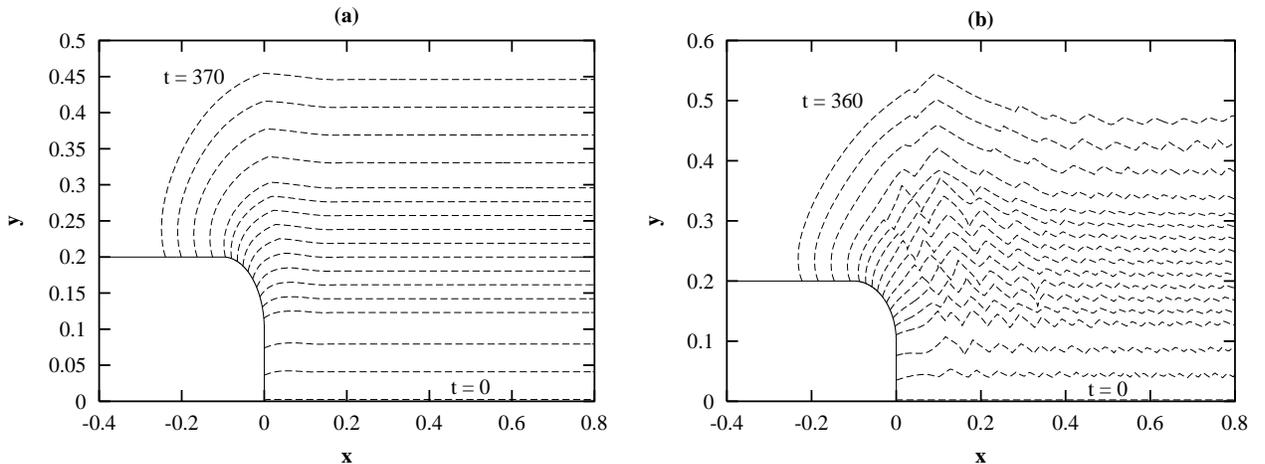,height=2.4in,width=6.5in,angle=270}
\caption{(a) The overgrowth with the strongly anisotropic interfacial energy over the blunt edge $r=0.5$ of the 
energetically inhomogeneous mask. $\Upsilon = 0.07$.  (b) Same as (a), $\Upsilon = 0.085$.}
\label{Fig8}
\end{figure}

Fig. \ref{Fig8} shows, as an example, the overgrowth over the blunt edge $r=0.5$
(the contact angle $\phi$ as in Fig. \ref{Fig3}(b)). In Fig. \ref{Fig8}(a), 
the value of the anisotropy 
parameter $\Upsilon$
(see \rf{2.13}) is 0.07. This is just above the critical value 1/15. $\Upsilon = 0.085$ in 
Fig. \ref{Fig8}(b). For both figures, the value of the phase angle $\beta_\gamma$ is chosen 1.67
(a little larger than $\pi/2$).

In Fig. \ref{Fig8}(a), the corner forms slowly at the elevation, and it is followed by the 
straight facet; other than that, the evolution resembles the isotropic evolution \cite{KBM1}.
In Fig. \ref{Fig8}(b), the interface instantly becomes unstable and the characteristic, coarsening 
with time hill-and-valley structure
is formed \cite{GDN,SGDNV}. As the result of coarsening, the single elevated corner
(bump) followed by the $\alpha=45^\circ$ 
inclined facet are again formed. Thus this numerical experiment suggests that the strong
interfacial energy anisotropy may be one of the possible causes for the ``bumpy" crystal shapes
routinely observed in ELO deposits \cite{THRUSH,Greenspan}.

\begin{figure}[H]
\centering
\psfig{figure=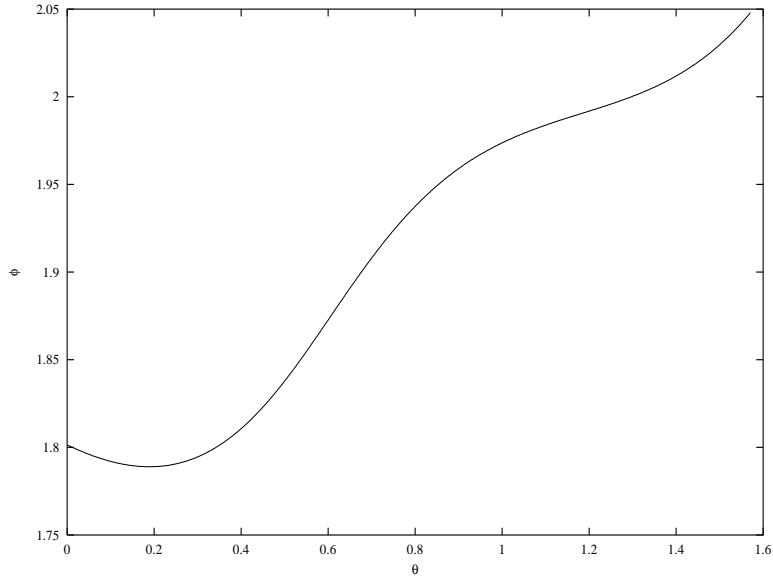,height=3.0in,width=4.0in,angle=270}
\caption{The contact angle \textit{vs.} $\theta$ for the interface evolution in Figures \ref{Fig8}(a,b).}
\label{Fig88}
\end{figure}

Fig. \ref{Fig88} shows the contact angle along the mask surface. In contrast to Figures \ref{Fig3}(b) and 
\ref{Fig5}(c), this function is not monotone. As $\epsilon_\gamma^{(0)}$ increases quite beyond value 0.01
chosen for computations in this section, the variations
in $\phi$ become more
pronounced and the computation becomes unstable.

\bigskip
\bigskip
\noindent
{\bf Acknowledgements}
\vspace{0.5cm}\\
\noindent
The author thanks Professors Richard Braun (University of Delaware) and Brian Spencer (SUNY at Buffalo) 
for the stimulating and useful discussions of this work. The reviewer is acknowledged for pointing out
the relevance of Herring's equation to the problem considered in this work.

\appendix
\renewcommand{\theequation}{\Alph{section}.\arabic{equation}}

\Section{On equation \rf{1.3}}

\begin{figure}[H]
\centering
\psfig{figure=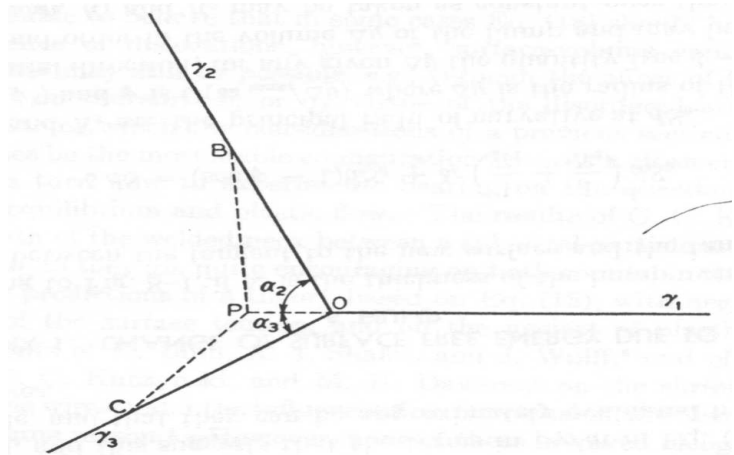,height=2.4in,width=3.8in,angle=0}
\caption{A sketch of the junction of three anisotropic interfaces, from \cite{Herring51}.}
\label{Fig99}
\end{figure}

The equation \rf{1.3} follows from the following equation derived by Herring \cite{Herring51}:
\begin{equation}
\gamma_1-\gamma_2\cos{\alpha_2} - \gamma_3\cos{\alpha_3} + \frac{\partial \gamma_2}{\partial \phi_2}\sin{\alpha_2}
+ \frac{\partial \gamma_3}{\partial \phi_3}\sin{\alpha_3} = 0.
\label{4.1}
\end{equation} 
The equation \rf{4.1} generalizes the Young-Dupr\'e equation 
\begin{equation}
\gamma_2\cos{\alpha_2} = \gamma_1-\gamma_3
\label{4.2}
\end{equation}
to orientational dependence of surface energies (tensions) at the tri-junction (see Fig. \ref{Fig99}). 
To obtain \rf{1.3} from \rf{4.1}, we identify the interface 1 in Fig. \ref{Fig99} with the mask-vapor interface,
the interface 2 with the crystal-vapor interface, the interface 3 with the crystal-mask interface and notice that
$\alpha_2=\phi$ and $\alpha_3=0$ since the slope of the mask surface is everywhere continuous (Fig. \ref{Fig2}). 

\renewcommand{\theequation}{\Beta{section}.\arabic{equation}}

\Section{Parameters of the model (characteristic of epitaxy of GaAs-like semiconductor at T$\approx$ 650$^\circ$C
)}

\bf{This appendix has been removed from online submission to arXiv.org in order to comply with
file size requirement; check the full version at\\ www.math.buffalo.edu/$\sim$mkhenner.}

\end{document}